\documentclass[11pt]{article}

\usepackage{mathtools}
\usepackage{amssymb}
\usepackage{dsfont}
\usepackage{slashed}
\usepackage{fullpage}
\usepackage{multirow}
\usepackage[colorlinks=true,linkcolor=blue,citecolor=magenta,linktocpage=true]{hyperref}
\usepackage{titlesec}
\titleformat*{\section}{\normalsize\bfseries}
\titleformat*{\subsection}{\normalsize\bfseries}
\titleformat*{\subsubsection}{\normalsize\bfseries}
\usepackage{cite}
\usepackage[most,breakable]{tcolorbox}
\usepackage{soul}
\setstcolor{red}
\usepackage{array}   % for \newcolumntype macro
\newcolumntype{C}{>{$}c<{$}}

\usepackage[numbers,sort,compress]{natbib}
\makeatletter
\renewcommand{\@dotsep}{10000}

\def\be#1\ee{\begin{align}#1\end{align}}
\def\bsub#1\esub{\begin{subequations}#1\end{subequations}}
\def\nn{\nonumber}
\def\q{\qquad}
\def\f{\frac}
\def\pe{\phantom{=}}

\def\ip{\lrcorner\,}
\def\ipp{\ip\!\!\!\ip}

\def\so{{\mathfrak{so}}}
\def\su{{\mathfrak{su}}}

\def\la{\langle}
\def\ra{\rangle}

\def\g{{\mathfrak{g}}}

\def\de{\mathrm{d}}
\def\d{\mathrm{d}}

\def\demi{{\frac12}}
\def\A{\mathcal{A}}
\def\cA{\mathcal{A}}
\def\cB{\mathcal{B}}
\def\B{\mathcal{B}}
\def\C{\mathcal{C}}
\def\cC{\mathcal{C}}
\def\D{\mathcal{D}}
\def\cD{\mathcal{D}}

\def\cF{\mathcal{F}}

\def\cG{\mathcal{G}}

\def\cJ{\mathcal{J}}
\def\J{\mathcal{J}}
\def\cJ{\mathcal{J}}

\def\L{\mathcal{L}}
\def\cL{\mathcal{L}}

\def\P{\mathcal{P}}

\def\cT{\mathcal{T}}
\def\T{\mathcal{T}}

\numberwithin{equation}{section}

\def\sdelta{\slashed \delta}

\begin{document}

\title{\Large{\textbf{\sffamily Diffeomorphisms as quadratic charges\\ in 4d BF theory and related TQFTs}}}
\author{\sffamily Marc Geiller$^1$, Florian Girelli$^2$, Christophe Goeller$^3$, Panagiotis Tsimiklis$^2$}
\date{\small{\textit{
$^1$Univ Lyon, ENS de Lyon, Univ Claude Bernard Lyon 1,\\ CNRS, Laboratoire de Physique, UMR 5672, F-69342 Lyon, France\\
$^2$Department of Applied Mathematics, University of Waterloo,\\ 200 University Avenue West, Waterloo, Ontario, Canada, N2L 3G1\\
$^3$Arnold Sommerfeld Center for Theoretical Physics,\\ Ludwig-Maximilians-Universit\"at M\"unchen,\\ Theresienstrasse 37, 80333 M\"unchen, Germany\\}}}

\maketitle

\begin{abstract}
We present a Sugawara-type construction for boundary charges in 4d BF theory and in a general family of related TQFTs. Starting from the underlying current Lie algebra of boundary symmetries, this gives rise to well-defined quadratic charges forming an algebra of vector fields. In the case of 3d BF theory (i.e. 3d gravity), it was shown in \cite{Geiller:2020okp} that this construction leads to a two-dimensional family of diffeomorphism charges which satisfy a certain modular duality. Here we show that adapting this construction to 4d BF theory first requires to split the underlying gauge algebra. Surprisingly, the space of well-defined quadratic generators can then be shown to be once again two-dimensional. In the case of tangential vector fields, this canonically endows 4d BF theory with a $\mathrm{diff}(S^2)\times\mathrm{diff}(S^2)$ or $\mathrm{diff}(S^2)\ltimes\mathrm{vect}(S^2)_\mathrm{ab}$ algebra of boundary symmetries depending on the gauge algebra. The prospect is to then understand how this can be reduced to a gravitational symmetry algebra by imposing Pleba\'nski simplicity constraints.
\end{abstract}

\thispagestyle{empty}
\newpage
\setcounter{page}{1}

\hrule
\tableofcontents
%\addtocontents{toc}{\protect\setcounter{tocdepth}{2}}
\vspace{0.7cm}
\hrule

%%%%%%%%%%%%%%%%%%%%%%%%%%%%%%%%%%%%%%%%%%%

\section{Introduction}

\subsection{Motivations}

Recently, a vast effort has been directed towards understanding the structure of gauge theories at codimension-2 boundaries. This is the location where symmetry charges and their algebras are revealed and may be studied in order to gain insights into the fine structure of the classical and quantum theory. In the context of gravity, this is of particular importance because (asymptotic) boundary charges provide the very definition of physical quantities such as energy and momenta, and encode a wealth of information about the interior region. This is relevant for the study of classical phenomena such as (say) radiation, but also holds promise for defining quantum gravity itself, for example through AdS/CFT and celestial (or flat space, or Carrollian) holography in the case of asymptotically-AdS and asymptotically-flat spacetimes.

It has been proposed recently that a notion of holography could be meaningful (and potentially more fundamental) even at finite distance \cite{Ciambelli:2021vnn,Ciambelli:2021nmv,Donnelly:2016auv,Speranza:2017gxd,Freidel:2020xyx,Freidel:2021yqe,Carrozza:2022xut}. Such a ``local holography'' is based on the study of the charges and symmetries associated with finite subregions. These come indeed equipped with a quasi-local algebra of charges, whose exact form depends on the formulation of gravity being considered and whose representation theory can in principle be used to define building blocks of the quantum theory. Since classical gravity contains this rich quasi-local symmetry structure and given the availability of geometric quantization tools \cite{Ciambelli:2022cfr,Donnelly:2020xgu,Barnich:2021dta}, it seems reasonable to try to approach quantum gravity from the complete understanding of the symmetry structure of the underlying (semi-)classical theory.

This picture, if taken seriously, suggests that one should aim at understanding the most general symmetry algebra which can be supported on boundaries of gravitational systems. In the case of asymptotic boundaries in three and four spacetime dimensions, this has in part motivated for example the study of various extensions of the BMS algebra. For arbitrary boundaries at finite distance, a universal notion of quasi-local boundary (or corner) symmetry algebra for metric gravity has been proposed in \cite{Ciambelli:2021vnn,Ciambelli:2021nmv} (see also \cite{Donnelly:2016auv,Speranza:2017gxd,Freidel:2020xyx,Freidel:2021yqe}). An important question which remains open is that of the dependency of the corner symmetry algebra on the variables and the formulation of gravity being considered. There are indeed general arguments showing that different formulations of gravity can turn on different corner charges (and potentially lead to a notion of dual charges \cite{Godazgar:2020gqd,Godazgar:2019dkh,Godazgar:2020kqd,Godazgar:2018qpq,Godazgar:2018dvh}) and thereby lead to different symmetry algebras.

An interesting setup to study these questions is 3-dimensional (3d hereafter) gravity. Since this theory is topological, one can reasonably think that a complete understanding of the boundary symmetry algebra should be achievable, both at finite distance and for asymptotic boundaries. It should also be possible to understand the differences (if any) between the metric and the connection-triad formulation. This has recently been studied in details in \cite{Geiller:2020okp,Geiller:2020edh}, where it has been shown that the obtention of diffeomorphism charges from charges of field-dependent internal gauge transformations naturally also leads to a notion of dual diffeomorphisms. This is the construction which we here aim at extending to 4-dimensional (4d) BF theory. For this, let us first recall the results of \cite{Geiller:2020okp,Geiller:2020edh}.

\subsection{Summary of the 3d construction}

The starting point of the analysis of \cite{Geiller:2020okp,Geiller:2020edh} is a general Lagrangian for 3d gravity, built from a triad $e$ and an independent spin connection $\omega$, and containing four Lorentz-invariant terms. This defines the so-called Mielke--Baekler model \cite{Mielke:1991nn}
\be\label{3d Lagrangian}
L=\f{\sigma_0}{3}e\wedge[e\wedge e]+2\sigma_1 e\wedge F+\sigma_2\, \omega\wedge\left(\de\omega+\f{1}{3}[\omega\wedge\omega]\right)+\sigma_3e\wedge\de_\omega e.
\ee
The couplings $(\sigma_0,\sigma_1,\sigma_2,\sigma_3)$ multiply respectively a volume term, the standard Hilbert--Palatini Lagrangian, the Chern--Simons Lagrangian for $\omega$, and a torsion term. Below we will also consider a general Lagrangian for 4d topological theories, and explore the role of the various couplings.

In order to understand the meaning of the couplings in \eqref{3d Lagrangian} we compute the equations of motion
\be\label{EOMs}
2F+p[e\wedge e]\approx0,
\q\q
2\de_\omega e+q[e\wedge e]\approx0,
\ee
which show that there is a source for internal curvature and torsion measured by the following combinations of the couplings of the Lagrangian:
\be
p\coloneqq\f{\sigma_0\sigma_1-\sigma_3^2}{\sigma_1^2-\sigma_2\sigma_3},
\q\q
q\coloneqq\f{\sigma_1\sigma_3-\sigma_0\sigma_2}{\sigma_1^2-\sigma_2\sigma_3}.
\ee
We restrict to the sector $\sigma_1^2-\sigma_2\sigma_3\neq0$ because otherwise the equations of motion only have the trivial solution $e=0$. The second equation of motion in \eqref{EOMs}, which is the torsion equation, is solved by $\omega=\Gamma-qe/2$ with $\Gamma(e)$ the torsionless Levi--Civita connection. Using this in the first equation of motion leads to its second order form, which is
\be
R_{\mu\nu}+2\lambda g_{\mu\nu}=0,
\q\q
\mathbb{R}\ni\lambda\coloneqq p+\f{q^2}{4}.
\ee
This metric equation of motion shows that the combination $\lambda$ of the initial four couplings of \eqref{3d Lagrangian} plays the role of the cosmological constant.

We are now interested in the symmetries and the charge algebra, which we want to study from two points of view: 1) the internal gauge transformations parametrized by Lie algebra elements, and 2) the diffeomorphism transformations parametrized by vector fields. First, the theory is invariant under the internal Lorentz transformations $\delta^\J$ and translations $\delta^\T$ acting as
\bsub\label{infinitesimal symmetries}
\be
\delta^\J_\alpha e&=[e,\alpha],&\delta^\T_\phi e&=\de_\omega\phi+q[e,\phi],\\
\delta^\J_\alpha\omega&=\de_\omega\alpha,&\delta^\T_\phi\omega&=p[e,\phi],
\ee
\esub
where $(\alpha,\phi)$ are Lie algebra-valued 0-forms. The covariant phase space formalism can then be used to derive the charges generating these Lorentz transformations and translations. For this we first compute the symplectic structure
\be
\Omega=\int_{M_2}\delta\theta,
\ee
where $\theta$ is the symplectic potential and $M_2$ is a 2d Cauchy slice with codimension-1 boundary $S$, and we then obtain the charges from the formula $\slashed{\delta}\J(\alpha)=-\delta^\J_\alpha\ipp\Omega$, where $\ipp$ denotes the interior product in field space. Restricting for the moment our attention to field-independent gauge parameters, we find that the charges are integrable and given by
\be\label{JT charges}
\J(\alpha)=2\oint_S\alpha(\sigma_1e+\sigma_2\omega),
\q\q
\T(\phi)=2\oint_S\phi(\sigma_1\omega+\sigma_3e).
\ee
We can then obtain the Poisson brackets of these charges from $\{\J(\alpha),\J(\beta)\}=-\delta^\J_\alpha\ipp\delta^\J_\beta\ipp\Omega$.
Defining the generator $\P\coloneqq\T-q\J/2$, we find that these charges form the centrally-extended current algebra
\bsub\label{JP current algebra}
\be
\{\J(\alpha),\P(\phi)\}&=\P([\alpha,\phi])-c_1\oint_S\alpha\de\phi,\\
\{\J(\alpha),\J(\beta)\}&=\J([\alpha,\beta])-c_2\oint_S\alpha\de\beta,\\
\{\P(\phi),\P(\chi)\}&=\lambda\left(\J([\phi,\chi])-c_2\oint_S\phi\de\chi\right),
\ee
\esub
where the central charges are $c_1=2\sigma_1-q\sigma_2$ and $c_2=2\sigma_2$. This is the current algebra built on the isometry algebra of the 3d Lorentzian spacetime with cosmological constant $\lambda$, i.e. $\mathfrak{iso}(2,1)$, $\mathfrak{so}(2,2)$, or $\mathfrak{so}(3,1)$ when $\lambda=0$, $\lambda>0$, and $\lambda<0$ respectively. This current algebra is defined at any location in the bulk of the theory, for any codimension-2 boundary. An interesting question is that of the status of the universal enveloping algebra, built from the generators \eqref{JT charges} by considering arbitrary field-dependent smearing gauge parameters $\alpha$ and $\phi$. A priori with arbitrary field-dependency the resulting charges are not integrable. This is the question we have studied in \cite{Geiller:2020okp,Geiller:2020edh} in the case of a field-dependency of the type\footnote{Here $\ip$ denotes the spacetime interior product.} $\xi\ip e$ and $\xi\ip\omega$, which gives rise to charges that are quadratic in the basic fields $(e,\omega)$. This is also what we aim at generalizing to 4d BF theory in the present work. As we will now explain, such quadratic charges are related to diffeomorphisms, and also give rise to a notion of dual diffeomorphisms.

Since the theory \eqref{3d Lagrangian} is topological, one can realize its diffeomorphisms on-shell of \eqref{EOMs} as field-dependent gauge transformations. Explicitly, one can check that the Lie derivative $\delta^\D_\xi=\L_\xi$ can be obtained as
\be\label{d=j+t}
\delta^\D_\xi\approx\delta^\J_{\xi\ip\omega}+\delta^\T_{\xi\ip e}.
\ee
We want to see what this entails at the level of the charges. Indeed, field-dependent gauge transformations do not generically lead to integrable charges unless one imposes extra conditions, on the dynamical fields and/or on the gauge parameters. Typically these conditions are automatically picked when building the residual symmetries for a spacetime with fixed (e.g. asymptotically-flat or AdS$_3$) boundary conditions. Here since we want to describe the symmetry charges associated to an arbitrary codimension-2 boundary we have to be careful about the integrability. One can easily check, either using the covariant phase space to construct the charge for the symmetry $\L_\xi$, or by computing $\slashed{\delta}\D(\xi)=\slashed{\delta}\J(\xi\ip\omega)+\slashed{\delta}\T(\xi\ip e)$, that when $\xi$ is field-independent the diffeomorphism charge is given by
\be\label{3d non-integrable diff}
\slashed{\delta}\D(\xi)\approx\oint_S\delta\big(2\sigma_1(\xi\ip\omega)e+\sigma_2(\xi\ip\omega)\omega+\sigma_3(\xi\ip e)e\big)-\xi\ip\theta,
\ee
where we have used $\slashed{\delta}$ to explicitly denote the fact that the charge is non-integrable. Indeed, it contains the familiar non-integrable piece $\xi\ip\theta$ for diffeomorphisms, where $\theta$ is the symplectic potential. As usual, this can be made integrable without boundary conditions on the fields by considering diffeomorphism transformations which do not move the boundary, i.e. tangent vector fields. For tangent vector fields, we then get the integrable charge
\be\label{usual diffeo 3d}
\D(\xi)=\J(\xi\ip\omega)+\T(\xi\ip e),
\ee
whose expression mirrors \eqref{d=j+t}.

The observation made in \cite{Geiller:2020okp,Geiller:2020edh} is that one can extend the present construction by asking the following question: Is there another combination of field-dependent internal gauge transformations which leads to integrable charges and to a closed algebra with the other symmetry generators? In other words, we view the current algebra \eqref{JP current algebra} as the fundamental algebra of the theory, and search for field-dependent generalizations. Since the generators $\J$ and $\P$ are linear in the fundamental fields $(e,\omega)$, searching for field-dependent combinations as in \eqref{usual diffeo 3d} amounts to considering quadratic generators. This is the spirit of the so-called Sugawara construction \cite{Sugawara:1967rw,DiFrancesco:639405,Caroca:2017onr}, which is usually written in terms of Fourier coefficients on the codimension-2 boundary (this interpretation will however not immediately be available in the 4d case later on). The answer to the above question is that there are indeed other field-dependent charges to consider, as the space of well-defined quadratics is actually 2-dimensional. More precisely, in addition to \eqref{usual diffeo 3d}, for tangent vector fields one may equally well consider the generator
\be\label{dual diffeo 3d}
\D^*(\xi)\coloneqq p\J(\xi\ip e)+\T(\xi\ip\omega)+\f{q}{2}\Big(\T(\xi\ip e)-\J(\xi\ip\omega)\Big).
\ee
Together with the above diffeomorphism charge this forms the algebra
\bsub\label{DD* algebra}
\be
\{\D(\xi),\D^*(\zeta)\}&=-\D^*([\xi,\zeta]),\\
\{\D(\xi),\D(\zeta)\}&=-\D([\xi,\zeta]),\\
\{\D^*(\xi),\D^*(\zeta)\}&=-\lambda\D([\xi,\zeta]).
\ee
\esub
Up to the central extensions, which are not present since here we are considering tangent vector fields, this is a ``diffeomorphism version'' of the $(\J,\P)$ current algebra \eqref{JP current algebra}. Starting from this latter, we have therefore managed to construct a well-defined algebra of quadratic generators whose parameters are vector fields. This algebra features once again the cosmological constant $\lambda$, and reduced in the case $\lambda=0$ to the centreless $\mathfrak{bms}_3$ algebra. When $\lambda\neq0$, we can alternatively introduce $\D^\pm\coloneqq(\D\pm\lambda^{-1/2}\D^*)/2$ to obtain the direct sum of two Witt (or centreless Virasoro) algebras.

It is clear from \eqref{DD* algebra} that the existence of $\D^*$ is due to the possibility of swapping the fields $e$ and $\omega$ in the field-dependent gauge parameters, which in turn is allowed since both fields are 1-forms. This immediately raises issues for a potential extension of this construction to the 4d case, where the fundamental fields are a 1-form and a 2-form. We will however solve this puzzle and show that the 4d theory also admits similar field-dependent charges in addition to the usual diffeomorphisms. Another insight into the duality in the 3d case comes from evaluating $\D$ and $\D^*$ on the asymptotic Killing vector field preserving e.g. BMS$_3$ boundary conditions. Doing so, one finds that $\D^*(\xi^u,\xi^r,\xi^\varphi)=\D(\xi^\varphi,\xi^r,\lambda\xi^u)$ \cite{Geiller:2020okp}, which shows that the duality is a modular duality between the temporal and angular directions. This is also a property which clearly cannot extend to a 4d setup since in this case there are two angular directions.

This brief summary of the construction of \cite{Geiller:2020okp,Geiller:2020edh} shows that there is a meaningful way in which one can realize a 2-dimensional algebra of vector fields, starting form the current algebra \eqref{JP current algebra}, by considering field-dependent internal gauge transformations. This algebra of vector fields reproduces, at finite distance and without the need for any boundary conditions, the centerless $\mathfrak{bms}_3$ or double Witt algebra depending on the cosmological constant. It should be noted that this generic construction cannot be achieved in the metric formulation (unless of course one introduces BMS or AdS$_3$ boundary conditions).

We now set out to build an analogue of this construction in 4d BF theory. Just like we have started here from a general 3d Lagrangian \eqref{3d Lagrangian}, in the 4d case we will also consider a general family of topological quantum field theories (TQFTs hereafter) in order to keep track of the role of the various coupling constants. Let us now turn to the 4d case.

\section{Quadratic charges in 4d BF theory}
\label{sec:4dBF1}

A similar structure to the one presented above in the 3d case exists for 4d TQFTs, with in particular a notion of dual diffeomorphisms which can be constructed from field-dependent internal gauge transformations. The construction is however not as straightforward as in the 3d case. In order to understand why this is so, let us first discuss the example of 4d BF theory.

The reason for considering 4d BF theory, especially with gauge group $\so(4)$ or $\so(3,1)$, is that it is closely related to Lorentzian or Euclidean 4d gravity (respectively) via the so-called Pleba\'nski formulation \cite{Buffenoir:2004vx,Plebanski:1977zz}, which is a rewriting of general relativity as a BF theory with constraints. This latter is at the basis of non-perturbative approaches to quantum gravity such as Loop Quantum Gravity and Spin Foams \cite{DePietri:1998mb,Perez:2004hj,Alexandrov:2011ab}, which have recently been revisited in the light of the developments on corner symmetries \cite{Freidel:2015gpa,Freidel:2016bxd,Freidel:2019ees,Freidel:2019ofr,Freidel:2020ayo,Freidel:2020xyx,Freidel:2020svx}. It was in particular explained in \cite{Freidel:2020ayo,Freidel:2020svx} how quantum numbers in the gravitational theory are inherited from the symmetries of BF theory. It will be interesting to understand in future work how this extends to the diffeomorphism symmetries constructed from quadratic charges. Indeed, one can hope that if a holographic dual can be constructed for 4d topological BF theory, then this boundary theory could be reduced to a gravitational boundary theory by implementing the Pleba\'nski simplicity constraints reducing BF theory to gravity.

\subsection{Summary of 4d BF theory}
\label{sec:4ddd?}

Let us consider as the basic variables a 2-form $\B$ and a connection 1-form $\A$, both valued in a Lie algebra $\g$. We will think of this algebra as being $\so(4)$ or $\so(3,1)$, although the details will not matter at this stage. The Lagrangian of 4d BF theory with quadratic potential is
\be\label{4d BF Lagrangian}
L=\la\cB \wedge \cF \ra - \kappa \la \cB \wedge \cB \ra,
\ee
where $\cF = \d\cA + \demi [\cA\wedge\cA]$ is the curvature of $\cA$. The pairing $\la \cdot \,, \cdot \ra: \g\times \g\to \mathbb{R}$ is given by the Killing form of $ \g$. We keep it explicit because it will play an important role in what follows. The equations of motion and the symplectic potential are
\be\label{4d BF EOMs}
\d_\cA \cB \equiv \d \cB +[\cA\wedge\cB]\approx0,
\q\q
\cF -2\kappa \cB\approx0,
\q\q
\theta =\la \cB\wedge \delta \cA \ra.
\ee
The symmetries of the Lagrangian \eqref{4d BF Lagrangian} are given by Lorentz transformations parametrized by $\g$-valued 0-forms $\alpha$, and translations parametrized by $\g$-valued 1-forms $\phi$. These internal gauge transformations act as
\bsub\label{4d BF infinitesimal symmetries}
\be
\delta^\J_\alpha\B&=[\B,\alpha],&\delta^\T_\phi\B&=\de_\A\phi,\\
\delta^\J_\alpha\A&=\de_\A\alpha,&\delta^\T_\phi\A&=2\kappa\phi.
\ee
\esub
As a remark, we note that computing the repeated action of these internal gauge transformations with field-independent parameters leads to
\be\label{global 4d BF algebra}
\big[\delta^\J_\alpha,\delta^\T_\phi\big]=\delta^\T_{[\alpha,\phi]},
\q\q
\big[\delta^\J_\alpha,\delta^\J_\beta\big]=\delta^\J_{[\alpha,\beta]},
\q\q
\big[\delta^\T_\phi,\delta^\T_\chi\big]=0,
\ee
meaning that the translations are abelian even for $\kappa\neq0$. This is in sharp contrast with the 3d theory \eqref{3d Lagrangian}, where even in the case $\sigma_2=0=\sigma_3$ the presence of a volume term $\sigma_0$ leads to non-abelian translations. The volume terms in \eqref{3d Lagrangian} and \eqref{4d BF Lagrangian}, with respective couplings $\sigma_0$ and $\kappa$, therefore play a different role from the point of view of the symmetries. We will come back to this observation later on.

As in the previous section, we can now use the covariant phase space to compute the charges generating these transformations. Restricting once again momentarily to field-independent gauge parameters, we find that the charges are given by
\be\label{initial BF charges}
\J(\alpha)=\oint_S\la\alpha,\B\ra,
\q\q
\T(\phi)=-\oint_S\la\phi\wedge\A\ra,
\ee
and that they form the centrally-extended current algebra
\bsub\label{4d BF JT current algebra}
\be
\{\J(\alpha),\T(\phi)\}&=\T([\alpha,\phi])-\oint_S\la\alpha,\de\phi\ra,\\
\{\J(\alpha),\J(\beta)\}&=\J([\alpha,\beta]),\\
\{\T(\phi),\T(\chi)\}&=2\kappa\oint_S\la\phi\wedge\chi\ra.
\ee
\esub
As expected, we recover the familiar fact that the algebra of charges is a centrally-extended current algebra based on the algebra of infinitesimal transformations \eqref{global 4d BF algebra}.

By analogy with the 3d construction introduced above, we now want to understand how diffeomorphisms are related to these charges of internal gauge transformations. Because \eqref{4d BF Lagrangian} is topological, we can once again realize the diffeomorphisms as field-dependent internal gauge transformations. This follows from the fact that
\bsub\label{4d BF onshell diffeo}
\be
\Big(\delta^\cJ_{\xi\ip \cA} +\delta_{\xi\ip \cB}^\cT \Big)\cA&= \cL_\xi \cA -\xi \ip (\cF-2\kappa \cB)\approx \delta^\cD_\xi \cA, \\
\Big(\delta^\cT_{\xi \ip \cB} + \delta^\cJ_{\xi\ip \cA} \Big)\cB& = \cL_\xi \cB -\xi \ip (\d_\cA \cB)\approx\delta^\cD_\xi \cB,
\ee
\esub
so as in the 3d case we have $\delta^\cD_\xi\approx\delta^\cJ_{\xi\ip \cA}+\delta_{\xi\ip \cB}^\cT$. For field-independent tangent vector fields to the (now) 2-dimensional boundary, we can therefore obtain an integrable diffeomorphism charge as the combination
\be\label{usual diffeo 4d}
\cD(\xi)= \cJ(\xi \ip \cA) + \cT(\xi \ip \cB).
\ee
While this formula is the 4d analogue of its 3d counterpart \eqref{usual diffeo 3d}, it is clear that the dual charge \eqref{dual diffeo 3d} defined above cannot be transposed to the 4d case since $\A$ and $\B$ are forms of different degrees. This is the immediate obstruction to defining an analogue of the dual diffeomorphisms in the 4d case.

We will now show that this roadblock is only apparent, and that it can be bypassed by performing a split of the fields $\A$ and $\B$ based on a split of the underlying algebra $\g$. When such a split is performed, a notion of duality becomes available and can in turn be used to define dual diffeomorphism charges.

Interestingly, this mechanism puts in a sense 4d BF theory on the same footing as 3d Chern--Simons (CS) theory. Indeed, if we consider 3d CS theory with $\g=\su(2)$, then there is only a single set of quadratic charges, corresponding to the usual diffeomorphisms. If, on the other hand, we consider CS theory where $\g$ is for example one of the isometry algebras of 3d gravity, then one can decompose the CS connection $\A$ into $e$ and $\omega$ using the decomposition of $\g$ (e.g. into rotations and boosts or rotations and translations depending on $\g$). The CS Lagrangian for $\A$ then becomes a Lagrangian for $e$ and $\omega$ as for example \eqref{3d Lagrangian} \cite{Geiller:2020edh}. As we have seen above, the possibility to consider the two independent quadratic charges \eqref{usual diffeo 3d} and \eqref{dual diffeo 3d} then arises since the new fields $e$ and $\omega$ can be swapped. The same mechanism is at play in 4d BF theory, where a decomposition of the algebra $\g$ induces a decomposition of both $\A$ and $\B$. Although there is no duality between $\A$ and $\B$ because of the mismatch in form degree, there is then a duality between the two components of $\A$ and a duality between the two components of $\B$. It is this duality which we want to explore and exploit in order to write down the independent quadratic charges. This is precisely what will lead to a 4d generalization of the ``dual'' diffeomorphisms built in 3d.

We first motivate and illustrate this construction on BF theory, and then present it in the case of a general 4d TQFT in order to mimic closely the structure of the 3d model \eqref{3d Lagrangian}.

\subsection{Construction of quadratic charges}

Let us now consider 4d BF theory with an algebra $\g$ whose generators can be decomposed into $(J_i,P_i)_{i=1,2,3}$ with brackets
\be\label{4d BF gauge structure}
[J_i,P_j] = \epsilon_{ij}{}^kP_k,
\q\q
[J_i,J_j] ={\epsilon_{ij}}^kJ_k,
\q\q
[P_i,P_j] =\lambda{\epsilon_{ij}}^kJ_k,
\ee
where $\lambda$ is a scalar parameter. Depending on the signature of $\epsilon$ (whose indices are lowered and raised with a Lorentzian or Euclidean $\eta$) and on the sign of $\lambda$, these commutation relations can be those of $\so(4)$, $\so(3,1)$, $\so(2,2)$, $\mathfrak{iso}(2,1)$, or $\mathfrak{iso}(3)$. We can then decompose the fields $\A$ and $\B$ in this basis as
\be\label{4d A and B decomposition}
\cA=A^i J_i+C^i P_i,
\q\q
\cB=B^i J_i+\Sigma^i P_i.
\ee
We also present in appendix \ref{app:self-dual basis} a construction of the quadratic charges using the self-dual basis. In order to decompose the Lagrangian in terms of these new fields, we need to pick a Killing form. There are two choices for this, given by
\bsub\label{4d BF pairings}
\begin{eqnarray}
\la P_i,J_j\ra_\text{I}=\eta_{ij}, &\q\q \la P_i,P_j \ra_\text{I}= 0, \q\q& \la J_i,J_j\ra_\text{I}= 0,\\
\la P_i,J_j\ra_\text{II}=0, & \la P_i,P_j \ra_\text{II}= \lambda \eta_{ij}, & \la J_i,J_j\ra_\text{II}= \eta_{ij}.
\end{eqnarray}
\esub
One should note that the second bilinear form is degenerate when $\lambda=0$. With these two choices, the Lagrangian \eqref{4d BF Lagrangian} becomes either of the two Lagrangians
\bsub\label{Lagrangians I and II}
\be
L_\text{I}&= \Sigma \wedge F- 2 \kappa B \wedge \Sigma+B \wedge \d_{A} C + \f{1}{2} \lambda \Sigma \wedge [C \wedge C],\label{Lagrangians I}\\
L_\text{II}&=B \wedge F -\kappa B \wedge B + \lambda \left(\Sigma \wedge \d_{A}C- \kappa  \Sigma \wedge \Sigma+ \f{1}{2}  B \wedge [C \wedge C] \right).\label{Lagrangians II}
\ee
\esub
In order to study these two Lagrangians in one go, we introduce below a general family of TQFTs described by \eqref{eq:monster}. The Lagrangians above are embedded in this family as
\be\label{I and II map to monster}
\begin{tabular}{|C|C|C|C|C|C|C|C|C|C|}
\hline
L_\text{gen} & ~\;\,\sigma_1\,\;~ & ~\;\,\sigma_2\,\;~ & ~\;\,\sigma_3\,\;~ & ~\;\,\sigma_4\,\;~ & ~\;\,\sigma_5\,\;~ & ~\;\,\sigma_6\,\;~ & ~\;\,\sigma_7\,\;~ & ~\;\,\sigma_8\,\;~ &  ~\;\,\sigma_9\,\;~  \\ \hline
L_\text{I} & 0 & 1 & 1 & 0 & 0 & 0 & -2\kappa & 0 & \lambda \\ \hline
L_\text{II} & 1 & 0 & 0 & \lambda & -2\kappa & -2\kappa\lambda & 0 & \lambda & 0 \\ \hline
\end{tabular}
\ee

When 4d BF theory is decomposed as in \eqref{Lagrangians I and II}, one has access to dualities which were not manifest in the initial Lagrangian \eqref{4d BF Lagrangian}. First, there is a duality between the 1-forms $A$ and $C$, and, second, there is a duality between the 2-forms $B$ and $\Sigma$. The presence of these dualities then allows to follow a construction similar to that present above in the 3d case, and to obtain quadratic charges which are independent from the ``usual'' diffeomorphisms. We are going to present the details of this construction in section \ref{sec:almostgeneral}, starting from the general family of 4d TQFTs described by \eqref{eq:monster}.

Before going into the details of this construction, let us collect the results in the case of 4d BF theory. The construction follows the same logic as in the 3d case. First, starting from \eqref{Lagrangians I} or \eqref{Lagrangians II}, one should find the internal gauge transformations, and then compute the associated charges and their algebra. This leads to the 4 charges \eqref{monster charges}, which we call $(\J_1,\J_2,\T_1,\T_2)$, and which correspond to a decomposition of the charges \eqref{initial BF charges} under \eqref{4d A and B decomposition}. Starting from these 4 charges, one can then build field-dependent charges by considering the contractions of the vector field $\xi$ with the fields $(A,C,B,\Sigma)$. In the case of the Lagrangians \eqref{Lagrangians I and II}, the analysis of \eqref{eq:monster} with the parameters \eqref{I and II map to monster} leads to 2 independent quadratic charges, which are moreover integrable when $\xi$ is tangent and field-independent. These charges are
\bsub\label{4d BF D and D*}
\be
\D(\xi)&=\cJ_1(\xi \ip A) + \cJ_2(\xi \ip C) + \cT_1(\xi \ip B) +  \cT_2 (\xi\ip \Sigma),\\
\D^*(\xi)&=\lambda\cJ_1(\xi\ip C) + \cJ_2 (\xi \ip A)+ \lambda \cT_1(\xi\ip \Sigma) + \cT_2(\xi \ip  B).
\ee
\esub
As announced, one can see that the essential difference between the new charges $\D^*$ and the usual diffeomorphisms $\D$ resides in the duality between the 1-forms and that between the 2-forms. Schematically, and up to the couplings, $\D^*$ is obtained from $\D$ by swapping the 1-forms $A$ and $C$ on the one hand, and the 2-forms $B$ and $\Sigma$ on the other hand. When $\xi$ is tangential and field-independent these charges are integrable and form the algebra
\bsub\label{4d BF D* and D algebra}
\be
\{\D(\xi),\D^*(\zeta)\} &= -\D^*([\xi,\zeta]),\\
\{\D(\xi),\D(\zeta)\} &= -\D([\xi,\zeta]),\\
\{\D^*(\xi),\D^*(\zeta)\} &= -\lambda\D([\xi,\zeta]).
\ee
\esub
One can therefore immediately see that this is a 4d generalization of the 3d algebra \eqref{DD* algebra}, where the vector fields $\xi$ now have components along the codimension-2 boundary $S$. The parameter $\lambda$ now controls the ``flat'' limit in which one of the two $\text{diff}(S^2)$ algebras becomes abelian. At the difference with the 3d case, where $\lambda$ is a combination of the initial couplings of the Lagrangian \eqref{3d Lagrangian}, in 4d BF theory $\lambda$ comes from the structure \eqref{4d BF gauge structure} of the gauge group, and the coupling of the Lagrangian \eqref{4d BF Lagrangian} does not appear in the quadratic charges nor in their algebra.

We now address the difference between the choices of pairings \eqref{4d BF pairings}, or equivalently the difference between the Lagrangians \eqref{Lagrangians I and II}. First, one should note that when $\lambda\neq0$ both Lagrangians in \eqref{Lagrangians I and II} give the same equations of motion, which are nothing but a decomposition using \eqref{4d A and B decomposition} of the equations of motion \eqref{4d BF EOMs}. Both Lagrangians also lead to the same algebra \eqref{4d BF D* and D algebra}. The choices I and II become however inequivalent when $\lambda=0$. Indeed, one can see that in this case the pairing II is degenerate. Because of this, in the limit $\lambda=0$ the Lagrangian $L_\text{II}$ degenerates, and reduces to that for a single ``half'' of the initial BF theory (for example $\su(2)$ if we start from $\so(4)$). More importantly, by carefully constructing the charges as we will do in the next section, one realizes that for $L_\text{II}$ the charges $\J_2$ and $\T_2$ are actually proportional to $\lambda$, so that in the limit $\lambda=0$ we get $\D^*=0$ altogether, and we are only left with $\D$ as the well-defined quadratic diffeomorphism charge. This echoes the remark which we have made at the end of section \ref{sec:4ddd?}.

Let us now turn to the detailed study of the general family of TQFTs in which the Lagrangians \eqref{Lagrangians I and II} are embedded. This will enable us to prove that the charges \eqref{4d BF D and D*} are indeed integrable and forming a closed algebra.

\section{Quadratic charges in a family of 4d TQFTs}
\label{sec:almostgeneral}

In this section, we introduce and study a family of 4d TQFTs which generalize, in a manner similar to \eqref{3d Lagrangian} in 3d, the Lagrangian for 4d BF theory, and which are also related to some specific examples of so-called 2-BF (or BFCG) theories \cite{Girelli:2007tt,Martins:2010ry,chen:2022}. We use this general TQFT to illustrate and present the details of the construction of the quadratic charges (and of the ``dual'' diffeomorphisms). At the difference with what we have presented above, here the starting point Lagrangian is written immediately in terms of the 4 fundamental fields: the connection 1-forms $A$ and $C$, and the 2-forms $B$ and $\Sigma$.
%The connection 1-forms $A$ and $C$ can respectively take value in $\g_1$ and $\g_2$, while the 2-forms $B$ and $\Sigma$ are respectively in $\g_1^*$ and $\g_2^*$. This therefore gives a generalization of the splitting \eqref{4d A and B decomposition}. We make the assumption that $\g_1$ and $\g_2$ have the same dimension (and hence also $\g_1^*$ and $\g_2^*$). We also consider that $\g_1$ is a Lie algebra, while for $\g_2$ this might not necessarily be the case. However, $\g_2$ is equipped with a bracket with value in $\g_1$ and/or $\g_2$. Finally, we have an action of $\g_1$ on $\g_2$ encoded in the bracket
%\be
%\g_1 \otimes \g_2 &\rightarrow \g_2\nonumber\\
%    X,Y &\rightarrow [X,Y].
%\ee
We use an index-free notation, the pairing is left implicit and does not play a role.

\subsection{Lagrangian and equations of motion}

Let us consider for the moment 9 arbitrary parameters $\sigma_{i}$, which we use as the couplings for 4 kinetic terms and 5 potential terms in order to build the Lagrangian
\be\label{eq:monster}
L_\text{gen}
&=\big(\sigma_1 B + \sigma_2 \Sigma\big) \wedge F + \big(\sigma_3 B + \sigma_4 \Sigma \big)\wedge \d_A C + \frac{\sigma_5}2 B \wedge B + \frac{\sigma_6}2 \Sigma \wedge \Sigma \cr
&\pe+\sigma_7 B \wedge \Sigma  + \left(\frac{\sigma_8}{2} B + \frac{\sigma_9}{2} \Sigma\right)\wedge [C\wedge C],
\ee
where $F = \d A + \demi [A\wedge A]$ and $\d_A C = \d C + [A\wedge C]$. One can see that when only $\sigma_1$ and $\sigma_5$ are non-vanishing we recover the standard BF Lagrangian with cosmological term. When only $\sigma_1$ and $\sigma_4=-\sigma_1$ are non-vanishing, we recover a specific 2-BF Lagrangian with 2-gauge symmetry \cite{Girelli:2007tt,Martins:2010ry,chen:2022}.

%defined in terms of the trivial/inner crossed module $(\g\stackrel{id}{\rightarrow} \g, \rhd)$, where $\rhd$ is the adjoint action  % \red{true for last point? or only particular case of bfcg ?}

When the 9 parameters $\sigma_{i}$ are all independent, the Lagrangian $L_\text{gen}$ does not yet describe a topological theory. We will explain below, based on a symmetry argument, that in order to obtain a topological theory there needs to be an algebraic relation between the $\sigma_i$'s. These couplings therefore give an 8-parameter family of TQFTs. We will come back to this shortly.

To obtain the symplectic potential and the equations of motion we now perform the variation of this Lagrangian, to find
\be\label{eq:variations}
    \delta L_\text{gen} = \d\theta
    &+ \delta B \wedge \left( \sigma_1 F +\sigma_3 \d_A C + \sigma_5 B + \sigma_7 \Sigma +\frac{\sigma_8}{2}[C\wedge C]\right) \cr
    &+ \delta \Sigma \wedge \left( \sigma_2 F + \sigma_4 \d_A C +\sigma_6 \Sigma +\sigma_7 B +\frac{\sigma_9}{2}[C\wedge C]\right) \cr
    &+ \delta A \wedge \Big ( \sigma_1 \d_A B + \sigma_2 \d_A \Sigma + \sigma_3 [C\wedge B] + \sigma_4 [C\wedge \Sigma] \Big) \cr
    &+ \delta C \wedge \Big( \sigma_3 \d_A B + \sigma_4 \d_A \Sigma + \sigma_8 [C\wedge B] + \sigma_9 [C\wedge \Sigma]\Big),
\ee
where the symplectic potential is
\be\label{monster potential}
\theta = \big(\sigma_1 B + \sigma_2\Sigma\big) \wedge \delta A + \big(\sigma_3 B + \sigma_4 \Sigma\big) \wedge \delta C.
\ee
We will use this symplectic potential below to compute the charges associated with the symmetries via the covariant phase space formalism. When $\sigma_1\sigma_4 - \sigma_2 \sigma_3 \neq 0$ (which is equivalent to the non-degeneracy condition for the 3d Lagrangian \eqref{3d Lagrangian}), the equations of motion can be rearranged in the form
\bsub
\be
      F &= p_1 B + q_1 \Sigma - \frac{r_1}{2} [C\wedge C], \\
      \d_A C &= p_2 B + q_2 \Sigma - \frac{r_2}{2}[C\wedge C], \label{eq:EOM_C}\\
      \d_A B &= p_3[B\wedge C]+q_3[\Sigma\wedge C],\\
      \d_A \Sigma &= p_4 [B\wedge C]+ q_3[\Sigma\wedge C]. \label{eq:EOM_Sigma}
\ee
\label{eq:EOM_tqft}
\esub
The new parameters appearing here are defined in terms of
\be
[ijkl] \coloneqq \frac{\sigma_i\sigma_j-\sigma_k\sigma_l}{\sigma_1\sigma_4 - \sigma_2\sigma_3}
\ee
by
\be
\begin{matrix}
p_1=[3745],&\quad&p_2=[2517],&\quad&p_3=[3428],&\quad&p_4=[1833]&\quad&r_1=[4839],\\
q_1=[3647],&\quad&q_2=[2716],&\quad&q_3=[4429],&\quad&q_4=[1934]&\quad&r_2=[1928].
\end{matrix}
\ee
The structure of the Lagrangian \eqref{eq:monster} and of its equations of motion already reveals that this is a 4d analogue, in the case of TQFTs, of the general 3d Lagrangian \eqref{3d Lagrangian}. Note that there could in principle be extra terms in \eqref{eq:monster} in order to symmetrize further the roles of $A$ and $C$ on the one-hand, and $B$ and $\Sigma$ on the other hand. These can however be obtained via field redefinitions.

Now that we have the Lagrangian that will serve as our starting point, we can proceed with the construction of quadratic charges following the 3d case. For this, the first step is to analyse the internal gauge symmetries and the resulting charge algebra.

\subsection{Symmetries, charges, and algebra}

We now turn to the study of the symmetries and the associated charges. Since we want the Lagrangian \eqref{eq:monster} to describe topological theories, we need to ensure that it has enough gauge symmetries to remove possible physical degrees of freedom\footnote{Essentially, in a topological theory there are enough symmetries to make any solution gauge equivalent to the trivial one.}. Furthermore, for a topological theory the diffeomorphisms can be written as field-dependent gauge transformations, so we need enough gauge transformations to realize this. In the example of BF theory studied in section \ref{sec:4dBF1}, there are two gauge symmetries, which are the Lorentz transformations and the translations. Here, since \eqref{eq:monster} has four fields one can expect that in the topological sector it admits four gauge symmetries. This is indeed the case provided we enforce a single algebraic constraint on the couplings $\sigma_i$.

More precisely, let us consider two Lie algebra-valued 0-forms $(\alpha,\chi)$ as well as two 1-forms $(\phi,\tau)$. If the couplings satisfy the constraint
\be\label{eq:topconstraint}
\sigma_1[7968]+\sigma_2[7859]+\sigma_3[3647]+\sigma_4[4537]\stackrel{!}{=}0,
\ee
then the Lagrangian \eqref{eq:monster} is invariant under the following gauge symmetries:
$$
\begin{array}{c|c|c|c}
\vphantom{\displaystyle\sum}  \text{0-form Lorentz}&\text{0-form translations}&\text{1-form translations}&\text{1-form Lorentz}\\\hline
\vphantom{\displaystyle\sum}  \delta^{\J_1}_\alpha A=\d_A\alpha&\delta^{\J_2}_\chi A=r_1[C,\chi]&\delta^{\T_1}_\phi A=p_1\phi&\delta^{\T_2}_\tau A=q_1\tau\\
\vphantom{\displaystyle\sum}  \delta^{\J_1}_\alpha C=[C,\alpha]&\delta^{\J_2}_\chi C=\d_A\chi+r_2[C,\chi]&\delta^{\T_1}_\phi C=p_2\phi&\delta^{\T_2}_\tau C=q_2\tau\\
\vphantom{\displaystyle\sum}  \delta^{\J_1}_\alpha B=[B,\alpha]&\delta^{\J_2}_\chi B=p_3[B,\chi]+q_3[\Sigma,\chi]&\delta^{\T_1}_\phi B=\d_A\phi+p_3[C\wedge\phi]&\delta^{\T_2}_\tau B=q_3[C\wedge\tau]\\
\vphantom{\displaystyle\sum}  \delta^{\J_1}_\alpha\Sigma=[\Sigma,\alpha]&\delta^{\J_2}_\chi\Sigma=p_4[B,\chi]+q_4[\Sigma,\chi]&\delta^{\T_1}_\phi\Sigma=p_4[C\wedge\phi]&\delta^{\T_2}_\tau\Sigma=\d_A\tau+q_4[C\wedge\tau]
  \end{array}
$$
This can be checked explicitly by plugging these symmetry transformations in the variational formula \eqref{eq:variations}. Of course, since any combination of these gauge symmetries is again a gauge symmetry, this parametrization is not unique. We have chosen it because it generalizes naturally the gauge structure of the so-called 2-BF theories studied in \cite{Girelli:2007tt,Martins:2010ry,chen:2022}.

Now that we have identified the symmetries of the Lagrangian, we can use the symplectic structure $\Omega$ derived from the potential \eqref{monster potential} in order to obtain the associated charges. Contracting each of the symmetry transformations found above in the symplectic structure, we find
\bsub\label{monster charges}
\be
    \delta \cJ_1(\alpha) =& -\int_{M_2} \alpha \; \delta \text{EOM}_A  + \oint_S \alpha (\sigma_1 \delta B +\sigma_2 \delta \Sigma),
    \\
    \delta \cJ_2(\chi) =& -\int_{M_2} \chi \; \delta \text{EOM}_C + \oint_S \chi  (\sigma_3 \delta B + \sigma_4 \delta \Sigma),
    \\
    \delta \cT_1(\phi) =& -\int_{M_2} \phi \wedge \delta \text{EOM}_B + \oint_S (\sigma_1\delta A +\sigma_3 \delta C )\wedge \phi,
    \\
    \delta \cT_2(\tau) =& -\int_{M_2} \tau \wedge \delta \text{EOM}_\Sigma  + \oint_S (\sigma_2 \delta A+\sigma_4 \delta C) \wedge \tau,
\ee
\esub
where $\text{EOM}_X$ refers to the equation of motion imposed by the field $X$ in \eqref{eq:variations}. From now on we will go on-shell of these equations of motion and consider the boundary charges. We will also consider for the moment field-independent gauge parameters, so these charges are manifestly integrable. Their algebra is found to be
\begin{subequations}\label{eq:pbgeneral}
  \begin{align}
    \{\J_1(\alpha),\J_1(\alpha')\} &= \J_1([\alpha,\alpha']),
    \\
    \{\J_1(\alpha),\J_2(\chi)\} &= \J_2([\alpha,\chi]) ,
    \\
    \{\J_2(\chi),\J_2(\chi')\} &= r_1\J_1([\chi,\chi']) +  r_2\J_2([\chi,\chi']),
    \\
    \{\J_1(\alpha),\T_1(\phi)\} &= \T_1([\alpha,\phi]) - \sigma_1 \oint_S \alpha \de\phi ,
    \\
    \{\J_1(\alpha),\T_2(\tau)\} &= \T_2([\alpha,\tau])-\sigma_2 \oint_S \alpha \de\tau ,
    \\
    \{\J_2(\chi),\T_1(\phi)\} &= p_3\T_1([\chi,\phi]) + p_4\T_2([\chi,\phi]) - \sigma_3 \oint_S \chi \de\phi ,
    \\
    \{\J_2(\chi),\T_2(\tau)\} &= q_3\T_1([\chi,\tau]) + q_4\T_2([\chi,\tau]) - \sigma_4 \oint_S \chi \de\tau ,
    \\
    \{\T_1(\phi),\T_1(\phi')\} &= -\sigma_5 \oint_S \phi \wedge \phi',\\
    \{\T_2(\tau),\T_2(\tau')\} &= -\sigma_6 \oint_S \tau \wedge \tau',
    \\
    \{\T_1(\phi),\T_2(\tau)\} &= -\sigma_7 \oint_S \phi \wedge \tau,
  \end{align}
\end{subequations}
and therefore contains 7 central extensions. In particular, one sees that the generators $(\T_1,\T_2)$ span a centrally-extended abelian current algebra, while the generators $(\J_1,\J_2)$ span a centreless $\so$-like current algebra. Defining the new generator $\widetilde{\J}_2\coloneqq \J_2-r_2\J_1/2$, one finds indeed that the subalgebra takes the form
\bsub\label{4d J1 J2 algebra}
\be
    \{\J_1(\alpha),\widetilde{\J}_2(\chi)\} &= \widetilde{\J}_2([\alpha,\chi]),\\
    \{\J_1(\alpha),\J_1(\alpha')\} &= \J_1([\alpha,\alpha']),\\
    \{\widetilde{\J}_2(\chi),\widetilde{\J}_2(\chi')\} &=\lambda  \J_1([\chi,\chi']) ,
\ee
\esub
with
\be\label{4d lambda}
\lambda \coloneqq r_1 + \f{r_2^2}{4}.
\ee
Interestingly, this is analogous to the structure \eqref{JP current algebra} which we have found in the 3-dimensional case, and it also represents a ``Lie algebra version'' of the diffeomorphism algebra which we are going to obtain below in the 4-dimensional case. Let us now explain how this algebra of diffeomorphisms is obtained from quadratic field-dependent charges.

\subsection{Field-dependent charges and (dual) diffeomorphisms}

The theory \eqref{eq:monster} is manifestly invariant under diffeomorphisms. We have also seen that when the couplings satisfy \eqref{eq:topconstraint} it describes a topological theory. As such, we expect that its diffeomorphisms can be rewritten on-shell as field-dependent gauge transformations, just like in the case of $d$-dimensional BF theory. One can explicitly check that this is indeed the case. More precisely, we find the natural analogues of \eqref{4d BF onshell diffeo}, given by the relations
\bsub
\be
\left( \delta_{\xi \ip A}^{\J_1} + \delta_{\xi \ip C }^{\J_2} + \delta_{\xi \ip B}^{\T_1} + \delta_{\xi \ip \Sigma}^{\T_2} \right) A
&=\L_\xi A -\xi \ip\left(F + \frac{r_1}2[C\wedge C]-p_1 B -q_1 \Sigma\right)\approx\delta^\D_\xi A, \\
\left( \delta_{\xi \ip A}^{\J_1} + \delta_{\xi \ip C }^{\J_2} + \delta_{\xi \ip B}^{\T_1} + \delta_{\xi \ip \Sigma}^{\T_2} \right) C
&=\L_\xi C-\xi \ip\left(\de_A C + \frac{r_2}2[C\wedge C] - p_2 B - q_2 \Sigma\right)\approx\delta^\D_\xi C, \\
\left( \delta_{\xi \ip A}^{\J_1} + \delta_{\xi \ip C }^{\J_2} + \delta_{\xi \ip B}^{\T_1} + \delta_{\xi \ip \Sigma}^{\T_2} \right) B
&=\L_\xi B-\xi \ip\Big(\de_A B -p_3 [B\wedge C] - q_3 [\Sigma\wedge C]\Big)\approx\delta^\D_\xi B, \\
\left( \delta_{\xi \ip A}^{\J_1} + \delta_{\xi \ip C }^{\J_2} + \delta_{\xi \ip B}^{\T_1} + \delta_{\xi \ip \Sigma}^{\T_2} \right) \Sigma
&=\L_\xi \Sigma-\xi \ip\Big(\de_A \Sigma - p_1 [B\wedge C] - q_1 [\Sigma\wedge C]\Big)\approx\delta^\D_\xi\Sigma ,
\ee
\esub
where $\cL_\xi (\cdot) = \d( \xi \ip (\cdot))+\xi \ip \d(\cdot)$.

To obtain the charges associated with diffeomorphism transformations in the topological sector of \eqref{eq:monster}, we can either directly contract the transformation $\delta^\D_\xi$ with the symplectic structure or, equivalently, use the field-dependent gauge transformations. In any case, we find that the charges are non-integrable, and that integrability can be achieved without boundary conditions on the dynamical fields if we restrict ourselves to field-independent tangential vector fields, as we did above when going from \eqref{3d non-integrable diff} to \eqref{usual diffeo 3d}. For such tangential vector fields, we then find that the integrable diffeomorphism charges are given in terms of the gauge charges \eqref{monster charges} by
\be\label{monster diffeos}
\cD(\xi) =  \cJ_1(\xi \ip A) + \cJ_2(\xi \ip C) + \cT_1(\xi \ip B) +  \cT_2 (\xi\ip \Sigma).
\ee
As expected, this expression is simply the analogue, in the case of the topological sector of \eqref{eq:monster}, of the relationship \eqref{usual diffeo 4d} obtained for 4d BF theory. We have indeed obtained the same formula \eqref{4d BF D* and D algebra} when decomposing the fields of 4d BF theory using \eqref{4d A and B decomposition}. We stress once again that here we are restricting ourselves to tangential vector fields in order to guarantee the integrability of the charges and so diffeomorphisms are span by a two-dimensional set of parameters.

Now, similarly to what was done in \cite{Geiller:2020edh} in the 3d case, we can show that the diffeomorphism \eqref{monster diffeos} is not the only integrable combination of quadratic field-dependent gauge transformations admitting a well-defined algebra with the gauge charges. There is an independent combination of field-dependent gauge charges which happens to be integrable in the case of field-independent tangential vector fields. This charge is given by
\be
\cC(\xi)\coloneqq\cJ_1\big(\xi\ip (r_1 C-p_3A)\big) +\cJ_2\big(\xi \ip (A+q_4C)\big)+ q_3 \cT_1(\xi\ip \Sigma) + \cT_2\big(\xi\ip(p_4 B+q_4\Sigma-p_3\Sigma)\big).
\label{eq:dual_charge}
\ee
We refer the reader to appendix \ref{ap:dual_diffeo_deriv} for the explicit proof of the integrability of this charge in the case of tangential $\xi$, and for the proof of the fact that its algebra with the charges $(\J_1,\J_2,\T_1,\T_2)$, and hence $\D$, is closed. This proof shows that the space of well-defined\footnote{Here by well-defined we mean that the charges i) are integrable in the case of tangential vector fields, and ii) have a closed algebra with the gauge charges $(\J_1,\J_2,\T_1,\T_2)$.} quadratic charges is 2-dimensional, and spanned by $\C$ and $\D$. One can note that the symmetry generated by this quadratic charge is
\be
\delta_\xi^\cC = \delta^{\cJ_1}_{\xi \ip (r_1C-p_3 A)} + \delta^{\cJ_2}_{\xi \ip(A+q_4C)} +q_3 \delta^{\cT_1}_{\xi \ip \Sigma}  + \delta^{\cT_2}_{\xi \ip (p_4 B+q_4\Sigma-p_3\Sigma)},
\ee
which gives the following infinitesimal transformations of the fields:
\bsub
\be
\delta^{\cC}_\xi A &=\cL_\xi\big(r_1C-p_3 A\big),\\
\delta^\cC_{\xi}C&= \cL_\xi\big(A + {q_4}C\big),\\
\delta_\xi^\cC B&= {q_3} \cL_\xi \Sigma,\\
\delta_\xi^\cC \Sigma &= \cL_\xi\big(p_4 B+q_4\Sigma-p_3\Sigma\big).
\ee
\esub
With all these ingredients, we can now compute the Poisson algebra between the two sets of quadratic charges. We find that this is given by
\bsub
\be
    \{ \cD(\xi),\cC(\zeta)\} &= -\cC([\xi,\zeta]),\\
    \{ \cD(\xi), \cD(\zeta)\} &=- \cD([\xi,\zeta]),\\
    \{ \cC(\xi),\cC(\zeta)\} &= (p_3-q_4)\cC([\xi,\zeta]) - p_4q_3\cD([\xi,\zeta]).
\ee
\esub
Introducing now the generator
\be\label{D* redefinition}
\cD^*(\xi)\coloneqq \C(\xi) +\f{1}{2}(p_3-q_4)\cD(\xi),
\ee
we find that the algebra takes the form
\bsub\label{final monster algebra}
\be
\{\cD(\xi),\cD^*(\zeta)\} &= -\cD^*([\xi,\zeta]),\\
\{ \cD(\xi), \cD(\zeta)\} &= - \cD([\xi,\zeta]),\\
\{ \cD^*(\xi), \cD^*(\zeta) \} &=  -\lambda \cD([\xi,\zeta]),
\ee
\esub
with $\lambda$ introduced above in \eqref{4d lambda}. This is a generalization to the case of the Lagrangian \eqref{eq:monster} of the diffeomorphism algebra \eqref{4d BF D* and D algebra} obtained above in 4d BF theory. One can also note that this diffeomorphism algebra is a ``vector field version'' of the current Lie algebra obtained above in \eqref{4d J1 J2 algebra}. We have therefore constructed a 4d generalization of the results of \cite{Geiller:2020edh} for the topological sector of the theory \eqref{eq:monster}.

One can note that these results are of course compatible with the ones obtained in \eqref{4d BF D* and D algebra} in section \ref{sec:4dBF1} for 4d BF theory, since plugging the couplings of table \eqref{I and II map to monster} into $r_1$ and $r_2$ leads indeed to $r_1+r_2^2/4=\lambda$, in agreement with the definition \eqref{4d lambda}.

The algebra \eqref{final monster algebra} features two commuting copies of a $\text{diff}(S^2)$ algebra. As in the 3d case, this can be made more explicit with the change of generators $\D^\pm\coloneqq(\D\pm\lambda^{-1/2}\D^*)/2$. We can also take the ``flat'' limit $\lambda\to0$ in order to obtain $\text{diff}(S^2)\ltimes\text{vect}(S^2)_\text{ab}$, which is a 4d generalization of $\mathfrak{bms}_3$ in the particular sense that $S^1$ is replaced by $S^2$. We recall that the unusual algebras \eqref{final monster algebra} for $\lambda\neq0$ and $\lambda=0$ have no gravitational interpretation, and have here been constructed as the algebras of tangential vector fields in a 4d topological theory. The novel feature, which extends the 3d results of \cite{Geiller:2020edh} to the 4d topological case, is that we have considered the quadratic charges in the current algebra \eqref{eq:pbgeneral} of the topological theory \eqref{eq:monster}. This shows that, when considering charges associated with finite regions in 4d BF theory, one can construct two diffeomorphism charges and obtain in a canonical manner the algebra \eqref{final monster algebra}.

\section{Perspectives}

In this paper we have proposed a Sugawara-type construction for a class of 4d TQFTs including 4d BF theory. This Sugawara construction starts from the current algebra of boundary symmetries, and then constructs quadratic generators forming a diffeomorphism algebra with vector fields as the parameters. More precisely, for 4d BF theory with an algebra \eqref{4d BF gauge structure}, starting from the current algebra \eqref{4d BF JT current algebra}, we have built the quadratic generators \eqref{4d BF D and D*} and shown that their algebra is \eqref{4d BF D* and D algebra}. In section \ref{sec:almostgeneral} we have then generalized this construction to the family of 4d TQFTs described by the Lagrangian \eqref{eq:monster}.

This construction extends to 4d the 3d construction presented in \cite{Geiller:2020okp,Geiller:2020edh}. As we have recalled, in 3d the existence of two independent sets of quadratic charges relies on a duality between the 1-forms $e$ and $\omega$. In the 4d case this duality is not available since the fields $\A$ and $\B$ of BF theory have a different form degree. However, we have argued that 4d BF theory is in fact the analogue of Chern--Simons theory in 3d, and that it is therefore natural to split its fields according to the decomposition \eqref{4d A and B decomposition} of the algebra \eqref{4d BF gauge structure}. When this split is performed, there is then a notion of duality within the 1-form and 2-form sectors, which allows to define the general quadratic charge \eqref{gdef}.

Surprisingly, as we have shown in the detailed proof of appendix \ref{ap:dual_diffeo_deriv}, for the examples of TQFTs described here the space of well-defined quadratic generators (which are integrable for tangential vector fields and which form a closed algebra) is 2-dimensional, just like in the 3d case. Although the quadratic generators $\D$ and $\D^*$ are not functionally independent from the generators $\J$ and $\T$ of the current algebra, they parametrize the independent symmetries of the theory. In 3d gravity in connection and triad variables, the diffeomorphisms are not the only gauge symmetries to which one can associate non-vanishing codimension-2 charges, since there are also the Lorentz transformations. Equivalently, one can replace these latter by the other set of (dual) diffeomorphisms, so as to preserve the size of the symmetry algebra. In 3d these extra quadratic generators have an interpretation in terms of a duality between the variables $(e,\omega)$, and also between the $u$ and $\phi$ directions in e.g. Bondi coordinates. Such an interpretation is not readily available in 4d, but should be investigated in future work, in particular to understand if the other set of quadratic charges (on top of the diffeomorphism charges) can be interpreted in terms of gravitational dual charges \cite{Godazgar:2020gqd,Godazgar:2019dkh,Godazgar:2020kqd,Godazgar:2018qpq,Godazgar:2018dvh}.

We should stress that it is not yet clear which condition should be satisfied by the internal gauge group in order for extra quadratic charges to exist. It is already clear from the construction of appendix \ref{app:self-dual basis} that one can exhibit BF theories for which the space of well-defined quadratic generators is more than 2-dimensional. One could hope for an algebraic criterion relating the properties of the gauge algebra to the size of the well-defined family of quadratic generators. Going even further, one can also wonder if the construction can be extended to cubic and higher order generators, so as to obtain $w_N$ algebras \cite{Lano:1992tc,KUPERSHMIDT_1993,artamonov2016introduction}. This is particularly interesting in light of the recent discussions on the $w_{1+\infty}$ symmetry present in (self-dual) gravity \cite{Adamo:2021lrv,Freidel:2021ytz,Ball:2021tmb,Strominger:2021lvk}. We should however stress that already in the case of 3d gravity the extension of the construction \cite{Geiller:2020okp} beyond second order has not yet been performed.

For future work, it would be interesting to investigate how the two sets of quadratic diffeomorphism charges relate to the bi-metric interpretation of BF theory \cite{Freidel:2008ku,Speziale:2010cf,Freidel:2012np}. It could indeed be the case that each quadratic diffeomorphism is a diffeomorphism charge for a specific metric sector of the bi-metric theory.

We also note that the construction presented here relies on a split of the internal gauge algebra, which essentially amounts to picking a preferred vector. In order to keep the internal covariance one should keep this vector as a variable and in particular include it in the covariant phase space analysis, as was done in \cite{Freidel:2020svx}. Just like in the case of gravity \cite{Freidel:2020ayo,Freidel:2020svx}, which as we recall can be obtained from BF theory by imposing simplicity constraints \cite{DePietri:1998mb,Perez:2004hj,Alexandrov:2011ab}, it would be interesting to perform the full covariant phase space analysis and the analysis of asymptotic symmetries for BF theory, either in terms of the bi-metric formulation or in terms of the bi-vector formulation. This would help understand the nature of the second set of quadratic charges (and in particular if they are ``dual'' charges as in the gravitational case \cite{Godazgar:2020gqd,Godazgar:2019dkh,Godazgar:2020kqd,Godazgar:2018qpq,Godazgar:2018dvh}). This could also help develop a boundary fluid interpretation for 4d topological theories such as BF theory \cite{Penna:2017vms,Penna:2017bdn,Donnelly:2020xgu}, and once again enable to reach gravitational results by imposing the simplicity constraints via the symmetry structure.

Finally, an interesting question is that of how to include central extensions in the algebra of quadratic charges by performing a higher-dimensional analogue of the twisted Sugawara construction which exists in 3d gravity (see e.g. appendix C.4 of \cite{Geiller:2020edh}).

\appendix

\section*{Acknowledgements}

FG thanks the Laboratoire de Physique at ENS de Lyon for hospitality during his stay as part of the ``Professeur Invité'' program. CG was supported by the Alexander von Humboldt Foundation.

\section{Derivation of the dual diffeomorphism charge}
\label{ap:dual_diffeo_deriv}

In this appendix we give the detailed derivation of the set of quadratic charges built from field-dependent generators of the current algebra. We require these quadratic charges to be integrable for tangential vector fields and to form a closed algebra with themselves and with the charges of the current algebra.

In order to perform this construction, let us consider the most general quadratic field-dependent combination of the charges $(\cJ_1,\cJ_2,\cT_1,\cT_2)$. This is given by
\be\label{gdef}
\sdelta \cG(\xi)=\sdelta\cJ_1\big(\xi\ip(aA+cC)\big)+\sdelta\cJ_2\big(\xi\ip(bC+dA)\big)+\sdelta\cT_1\big(\xi\ip(eB+g\Sigma)\big)+\sdelta\cT_2\big(\xi\ip(f\Sigma+hB)\big),
\ee
where the parameters $(a,b,c,d,e,f,g,h)$ are to be determined and the vector field $\xi$ is field-independent. We now impose the following two requirements on this general quadratic charge:
\begin{itemize}
  \item[1)] The charges must be integrable when the vector fields $\xi$ are tangent to the codimension-2 surface $S$.
  \item[2)] The charges must form a closed algebra (up to possible central extensions) with the gauge charges \eqref{monster charges} of the initial current algebra.
\end{itemize}

Let us start with the condition of integrability. Using the explicit form of the gauge charges, we find that $\sdelta\cG(\xi)$ takes the form
\bsub
\be
\sdelta\cG(\xi)
&=\xi \ip (a A +c  C) (\sigma_1 \delta B +\sigma_2 \delta \Sigma) + \xi\ip (b C + d A) (\sigma_3 \delta B +\sigma_4 \delta \Sigma)\cr
&\pe+(\sigma_1 \delta A + \sigma_3 \delta C) \wedge \xi \ip (e B+g\Sigma) + (\sigma_2 \delta A + \sigma_4 \delta C)\wedge \xi\ip (f\Sigma+h B)\label{eq:cG_explicit_form}\\
&=\delta \Big( \xi\ip (aA+cC) \wedge (\sigma_1 B +\sigma_2 \Sigma) +  \xi\ip (bC+dA)\wedge (\sigma_3 B +\sigma_4 \Sigma) \Big)\cr
&\pe-\xi \ip \Big( (a\delta A+ c\delta C)\wedge (\sigma_1 B + \sigma_2 \Sigma) + (b \delta C+d \delta A)\wedge (\sigma_3 B + \sigma_4 \Sigma) \Big)\cr
&\pe-(\sigma_1a +\sigma_3d - \sigma_2 h -\sigma_1 e) \delta A \wedge \xi \ip B  -(\sigma_2 a +\sigma_4 d -\sigma_1 g -\sigma_2 f) \delta A\wedge \xi\ip \Sigma\cr
&\pe-(\sigma_1 c +\sigma_3 b -\sigma_3 e -\sigma_4 h) \delta C \wedge \xi\ip B -(\sigma_2 c +\sigma_4 b -\sigma_3 g -\sigma_4 f) \delta C \wedge \xi \ip \Sigma.\label{eq:cG_rewritten_form}
\ee
\esub
Equation \eqref{eq:cG_explicit_form} is the explicit expression for $\sdelta\cG(\xi)$ in terms of the gauge charges \eqref{monster charges} of the current algebra. Equation \eqref{eq:cG_rewritten_form} is then a rewriting of $\sdelta\cG(\xi)$ which isolates the integrable part and the part which vanishes when $\xi$ is tangent to $S$. The last two lines are four independent and generically non-integrable contributions to $\sdelta\cG(\xi)$. To obtain \eqref{eq:cG_rewritten_form}, we performed a $\delta$ integration by part over the first line of \eqref{eq:cG_explicit_form}, followed by an ``integration by parts'' on $\xi\ip$ in the non-exact term arising from the $\delta$ integration by part. Note that this procedure is of course not unique. Instead of using the first line of \eqref{eq:cG_explicit_form}, we could have used the second one. While this choice does affect the intermediate results, the final result of this appendix is of course independent of it.

Without additional conditions on the fields and/or on the vector field $\xi$ to render the last two lines of \eqref{eq:cG_rewritten_form} integrable, we need to enforce the following four conditions on the parameters in order for $\sdelta \cG(\xi)$ to be integrable:
\bsub\label{eq:integrability_constraint}
\be
    \sigma_1 a +\sigma_3d - \sigma_2 h -\sigma_1 e &= 0,\\
    \sigma_2 a +\sigma_4 d -\sigma_1 g -\sigma_2 f&=0 ,\\
    \sigma_1 c +\sigma_3 b  -\sigma_4 h-\sigma_3 e&=0,\\
    \sigma_2 c +\sigma_4 b -\sigma_3 g -\sigma_4 f&=0.
\ee
\esub
We now assume that $(a,b,c,d,e,f,g)$ are such that the constraints \eqref{eq:integrability_constraint} hold, and we therefore consider the 4-dimensional space (spanned by $(a,b,c,d,e,f,g)$ respecting \eqref{eq:integrability_constraint}) of integrable charges of the form
\be
\cG(\xi)=\cJ_1\big(\xi\ip (a A + c C)\big) + \cJ_2 \big(\xi\ip (b C + d A)\big) + \cT_1\big(\xi\ip (e B + g \Sigma)\big) + \cT_2\big(\xi\ip (f\Sigma + h B)\big) .
\ee

We now look at the Poisson brackets of these charges with the charges $(\cJ_1,\cJ_2,\cT_1,\cT_2)$ of the initial current algebra. We first consider the bracket with $\cT_1(\phi)$, which reads
\be
  \{\cG(\xi) , \cT_1(\phi) \}
  &=
  \cT_1\big([\xi\ip (aA+cC),\phi]\big) -\sigma_1 \oint_S \xi\ip (aA+cC)\wedge \d \phi\cr
  &\pe+p_3 \cT_1\big([\xi\ip(dA+bC),\phi]\big)+p_4 \cT_2\big([\xi\ip(dA+bC),\phi]\big) -\sigma_3\oint_S   \xi\ip(dA+bC)\wedge \d\phi \cr
  & \pe-\sigma_5 \oint_S  \xi\ip(eB+h\Sigma)\wedge \phi \cr
  & \pe- \sigma_7 \oint_S \xi\ip(gB+f\Sigma)\wedge \phi.
\ee
Here the first line corresponds to the Poisson bracket between $\cJ_1\big(\xi\ip (a A + c C)\big)$ and $\cT_1(\phi)$, the second between $ \cJ_2 \big(\xi\ip (b C + d A)\big)$ and $\cT_1(\phi)$, the third between $\cT_1\big(\xi\ip (e B + g \Sigma)\big)$ and $\cT_1(\phi)$ and the fourth line between $\cT_2\big(\xi\ip (f\Sigma + h B)\big)$ and $\cT_1(\phi)$. To go further, we now use the explicit expressions for the charges $\cT_1$ and $\cT_2$ to write
\be
  \{\cG(\xi) , \cT_1(\phi) \}
&=
  \oint_S \big( \sigma_1 a + (p_3 \sigma_1 + p_4\sigma_2)d \big)[A,\xi \ip A] \wedge \phi + \big( \sigma_3 c + (p_3\sigma_3 + p_4\sigma_4) b\big)[C,\xi\ip C]\wedge \phi \nn\\
  &\pe+\oint_S \big( \sigma_1 c + (p_3\sigma_1 +p_4\sigma_2) b \big) [A,\xi \ip C]\wedge \phi + \big( \sigma_3a + (p_3\sigma_3  + p_4 \sigma_4) d)  \big)[C,\xi\ip A] \wedge \phi \nn\\
  &\pe-\sigma_1 \oint_S  \big(\xi\ip (a \d A+c \d C)\big) \wedge \phi
  -\sigma_3\oint_S  \big(\xi\ip(d \d A+b \d C)\big) \wedge \phi \nn\\
  &\pe-\sigma_5 \oint_S \big(\xi\ip(eB+h\Sigma)\big)\wedge \phi - \sigma_7 \oint_S \big(\xi\ip(gB+f\Sigma)\big)\wedge \phi,
\ee
where we have integrated by parts on $\d\phi$ and used the relation $[P \wedge Q] \wedge R = (-1)^{(p+q)r} [R \wedge P] \wedge Q$ in order to isolate $\phi$. Now, noting that the couplings satisfy the relations
\be
  p_3 \sigma_1 + p_4 \sigma_2 = \sigma_3, \q\q p_3 \sigma_3 + p_4 \sigma_4 = \sigma_8,
\ee
one can rewrite the bracket as
\be
  \{\cG(\xi) , \cT_1(\phi) \}
  &=\oint_S \big( \sigma_1 a + \sigma_3 d \big)[A,\xi \ip A] \wedge \phi + \big( \sigma_3 c + \sigma_8 b\big)[C,\xi\ip C]\wedge \phi \cr
  &\pe+\oint_S \big( \sigma_1 c + \sigma_3 b \big) [A,\xi \ip C]\wedge \phi + \big( \sigma_3a + \sigma_8 d)  \big)[C,\xi\ip A] \wedge \phi \cr
  &\pe-\sigma_1 \oint_S  \big(\xi\ip (a \d A+c \d C)\big) \wedge \phi
  -\sigma_3\oint_S  \big(\xi\ip(d \d A+b \d C)\big) \wedge \phi \cr
  &\pe-\sigma_5 \oint_S \big(\xi\ip(eB+h\Sigma)\big)\wedge \phi - \sigma_7 \oint_S \big(\xi\ip(gB+f\Sigma)\big)\wedge \phi.
\ee
Finally, using Cartan's magic formula, the fact that the vector field $\xi$ is tangential, and massaging the various terms, the bracket becomes
\be\label{eq:bracket_G_T1_pre}
   \{\cG(\xi), \cT_1(\phi)\} &=
    b_1 \cT_1(\cL_\xi \phi) +b_2 \cT_2(\cL_\xi \phi) \cr
    &\pe- \oint_S \xi \ip \Big( ( \sigma_1 a + \sigma_3 d) F + \f{1}{2}(\sigma_3 c + \sigma_8 b)[C \wedge C] + ( \sigma_1 c +\sigma_3 b ) \d_A C  \cr
    &\pe \q\q+ (\sigma_5 e + \sigma_7 g) B + (\sigma_5 h + \sigma_7 f) \Sigma \Big) \wedge \phi \cr
    &\pe- \oint_S (\sigma_1 c +\sigma_3 b-\sigma_3a -\sigma_8 d  ) [\xi\ip A, C]\wedge \phi,
\ee
where
\bsub
\be
b_1= \frac{\sigma_1 \sigma_2 c+\sigma_2 \sigma_3 b-\sigma_1\sigma_4 a -\sigma_3\sigma_4 d}{\sigma_1\sigma_4-\sigma_2 \sigma_3},
\q\q
b_2= \frac{\sigma_1 \sigma_3 a +\sigma_3^2 d -\sigma_1^2 c - \sigma_1 \sigma_3 b}{\sigma_1\sigma_4-\sigma_2 \sigma_3}.
\ee
\esub
One can see in \eqref{eq:bracket_G_T1_pre} that the Poisson bracket of $\cG$ with $\cT_1$ has two types of contributions. The terms on the first line are once again charges of the current algebra. The last three lines however are neither expressible in terms of gauge charges nor central extensions. Therefore, the only possibility for the algebra to close is that somehow these terms cancel or vanish on-shell. More precisely, we would like to find a condition on the parameters such that the bracket takes the form
\be\label{condition bracket eom}
\{\cG(\xi), \cT_1(\phi)\} \stackrel{?}{=}b_1 \cT_1(\cL_\xi \phi) +b_2 \cT_2(\cL_\xi \phi)-\oint_S \xi \ip \Big( \alpha \text{EOM}_B + \beta p_4 \text{EOM}_\Sigma \Big) \wedge \phi,
\ee
where $(\alpha,\beta)$ are arbitrary parameters and where the rescaling by $p_4$ has been introduced for later convenience. The reason for which only the equations of motion enforced by $B$ and $\Sigma$ appear here is because the terms which need to be cancelled in the bracket \eqref{eq:bracket_G_T1_pre} only involve contributions from these equations of motion. It is clear that the last line in \eqref{eq:bracket_G_T1_pre} never appears in the equations of motion. We therefore need to impose that this term is vanishing, which amounts to the condition
\be\label{eq:condition_bracket_1}
  \sigma_1 c +\sigma_3 b = \sigma_3 a +\sigma_8 d .
\ee
Then a comparison with the equations of motion \eqref{eq:EOM_tqft} tells us that the remaining terms in \eqref{eq:bracket_G_T1_pre} can be written in the form appearing in \eqref{condition bracket eom} provided we have the following relations:
\bsub\label{eq:condition_bracket_2}
\be
  \alpha\sigma_1 +\beta p_4 \sigma_2 =& \sigma_1a +\sigma_3 d , \\
  \alpha\sigma_8 +\beta p_4 \sigma_9 =& \sigma_3 c +\sigma_8 b ,\\
  \alpha \sigma_3 +\beta p_4 \sigma_4 =& \sigma_1 c +\sigma_3 b ,\\
  \alpha \sigma_5 +\beta p_4 \sigma_7 =& \sigma_5 e +\sigma_7 g , \label{ge}\\
  \alpha \sigma_7 +\beta p_4 \sigma_6 =& \sigma_5 h + \sigma_7 f .\label{hf}
\ee
\esub
In summary, the coupling therefore have to satisfy the 10 equations \eqref{eq:integrability_constraint}, \eqref{eq:condition_bracket_1}, and \eqref{eq:condition_bracket_2}. This system is overdetermined, but luckily some of these equations are redundant. The system can be solved and the space of solutions is in fact 2-dimensional. It can be easily parametrized by two parameters $(x,y)$, in terms of which we get
\bsub
\be
    a &= x -y p_3 ,\\
    b &= x +y q_4 ,\\
    c &= y r_1 ,\\
    d &= y ,\\
    e &= x  ,\\
    f&= x+ y(q_4-p_3) ,\\
    g &= y q_3 ,\\
    h &= y p_4,
\ee
\esub
provided that the $\sigma$'s satisfy
\be
    \sigma_5 q_3 +\sigma_7 q_4 -\sigma_7 p_3 -\sigma_6 p_4 =0,
\ee
which is in fact equivalent to the condition \eqref{eq:topconstraint} which needs to hold in order for the theory to be topological.

We have therefore found a 2-dimensional space of parameters within $(a,b,c,d,e,f,g,h)$ which gives integrable charges $\cG(\xi)$ satisfying a closed algebra with $\cT_1(\phi)$. Remarkably, the conditions on the parameters also guarantee that the other Poisson brackets are closed. If we denote a quadratic charge which solve the above conditions by $\cG_{x,y}(\xi)$, where $(x,y)$ are the solutions of the above systems, then the Poisson brackets with the gauge charges are
\bsub
\be
    \{\cG_{x,y}(\xi), \cT_1(\phi)\} &= -x \cT_1 (\cL_\xi \phi) -y p_4 \cT_2 (\cL_\xi \phi) ,\\
    \{\cG_{x,y}(\xi), \cT_2(\tau)\} &= -y q_3\cT_1(\cL_\xi \tau) - \big(x +y(q_4-p_3)\big)\cT_2(\cL_\xi \tau),\\
    \{ \cG_{x,y}(\xi), \cJ_1(\alpha)\} &= -\left(x - y p_3\right) \cJ_1(\cL_\xi \alpha) -y \cJ_2(\cL_\xi \alpha),\\
    \{\cG_{x,y}(\xi), \cJ_2(\chi)\} &=-y r_1\cJ_1(\cL_\xi\chi) - (x +y q_4)\cJ_2(\cL_\xi \chi).
\ee
\esub
We note that the brackets with $\cJ_1$ and $\cJ_2$ are closed regardless of the constraint on the parameters.

We can now pick two convenient representatives in the 2-parameter family $\cG_{x,y}$ of quadratic charges. The usual diffeomorphism corresponds to taking $(x,y)=(1,0)$, for which we find
\be
  \cG_{1,0}(\xi)=\cD(\xi) = \cJ_1(\xi\ip A) +  \cJ_2 (\xi\ip C) + \cT_1(\xi\ip B) +\cT_2(\xi\ip \Sigma) .
\ee
An obvious other independent quadratic charge is found by taking $(x,y)=(0,1)$. In this case we obtain the charge \eqref{eq:dual_charge} mentioned in the main text, i.e.
\be
  \cG_{0,1}(\xi) = \cC(\xi) = \cJ_1\big(\xi\ip (r_1 C-p_3A)\big) +\cJ_2\big(\xi \ip (A+q_4C)\big)+ q_3 \cT_1(\xi\ip \Sigma) + \cT_2\big(\xi\ip(p_4 B+q_4\Sigma-p_3\Sigma)\big).
\ee
These two charges $(\cD,\cC)$ form a basis of the integrable quadratic charges for tangent vector fields, and we have
\be
  \cG_{x,y}(\xi) = x \cD(\xi) + y \cC(\xi) .
\ee

Finally, we want to compute the bracket between these quadratic charges themselves. To do so, we can either use the definition of the quadratic charges and the elementary brackets of the current algebra, or alternatively use the action of the quadratic generators on the fields. By construction, the diffeomorphism acts by the Lie derivative. On the other hand, $\cC$ acts as
\bsub
\be
    \delta^{\cC}_\xi A &=\cL_\xi\big(r_1C-p_3 A\big),\\
    \delta^\cC_{\xi}C&= \cL_\xi\big(A + {q_4}C\big),\\
    \delta_\xi^\cC B&= {q_3} \cL_\xi \Sigma,\\
    \delta_\xi^\cC \Sigma &= \cL_\xi\big(p_4 B+q_4\Sigma-p_3\Sigma\big).
\ee
\esub
Using these expressions in the covariant phase space formula for the Poisson brackets, it can be shown that the algebra between $(\cD,\cC)$ is closed and takes the form
\bsub
\be
    \{ \cD(\xi),\cC(\zeta)\} &= -\cC([\xi,\zeta]) ,\\
    \{ \cD(\xi), \cD(\zeta)\} &=- \cD([\xi,\zeta])  ,\\
    \{ \cC(\xi),\cC(\zeta)\} &= (p_3-q_4)\cC([\xi,\zeta]) - p_4q_3\cD([\xi,\zeta]) ,
\ee
\esub
which can further be rewritten as \eqref{final monster algebra} upon redefining the new generator \eqref{D* redefinition}.

\section{Self-dual basis and additional quadratic charges}
\label{app:self-dual basis}

In this appendix, we present another derivation of the two sets of quadratic charges in 4d BF theory. While this derivation is less general than the one presented in the main text, in the sense that it requires an additional property on the underlying algebra, it allows to understand possible generalizations to the construction of the admissible quadratic charges. We illustrate this at the end of the appendix on the example of BF theory with gauge algebra $\g = \su(2) \oplus \su(2) \oplus \su(2)$.

In the following we focus on the case of $\g =\so(4)$ (the generalization to $\so(3,1)$ or $\so(2,2)$ can easily be done). Recall that the algebra then has the form \eqref{4d BF gauge structure}
\be
  [J_i,P_j] = \epsilon_{ij}{}^kP_k,
\q\q
  [J_i,J_j] ={\epsilon_{ij}}^kJ_k,
\q\q
  [P_i,P_j] =\lambda {\epsilon_{ij}}^kJ_k,
\ee
where $\lambda$ is never vanishing. This is the Cartan decomposition. These exists however another decomposition of these algebra, the so-called self-dual decomposition, which allows to have another viewpoint on the existence of multiple quadratic charges. The self-dual basis is induced by the following change of basis\footnote{The notation $1,2$ instead of the standard $\pm$ will become clearer at the end of this appendix, when we consider the case with more subalgebras in $\g$.}
\begin{equation}
  \sigma^1_i = J_i +\sqrt{ \lambda}\, P_i,\q\q \sigma^2_i = J_i - \sqrt{ \lambda} \,P_i,
  \label{eq:change_basis}
\end{equation}
such that the brackets become
\begin{equation}
  [\sigma_i^a,\sigma^b_j] =2 \delta^{ab} \epsilon_{ij}{}^k\sigma^{a}_k .
\end{equation}
The self-dual basis therefore splits the algebra into two commuting subalgebras. For example, in the case of $\so(4)$, each subalgebra is isomorphic to $\su(2)$. Note that this construction only works if $\lambda$ is non-vanishing, which is why we are imposing this condition. The fields $\cA$ and $\cB$ can be decomposed in either of the above basis, and we have
\begin{subequations}
  \begin{equation}
    \cA=A^i J_i +C^i P_i  = A_1^i\sigma^1_i+A_2^i\sigma_i^2, \q A_1^i = \demi(A^i+ C^i), \q A_2^i = \frac{1}{2}(A^i- C^i) \; ,
  \end{equation}
  \begin{equation}
    \cB=B^i J_i +\Sigma^i P_i  = B_1^i\sigma^1_i+B_2^i\sigma_i^2, \q  B_1^i = \demi(B^i+ \Sigma^i), \q B_2^i = \frac{1}{2}(B^i- \Sigma^i).
  \end{equation}
\end{subequations}
The main advantage of the self-dual basis is that it allows to diagonalize the Killing forms of the algebra
\begin{align}
    \left.\begin{array}{c}
    \left(\begin{array}{l}
    \la P_i,J_j\ra =0 \\ \la P_i,P_j \ra = \lambda \eta_{ij} \\ \la J_i,J_j\ra = \eta_{i,j}
    \end{array} \right) \\
    \left(\begin{array}{c}
    \la P_i,J_j\ra=\eta_{ij}\\ \la P_i,P_j\ra = 0 =\la J_i,J_j\ra
    \end{array} \right)
    \end{array} \right\}
    \quad   \longrightarrow \quad
    \left\{\begin{array}{c}
    \left(\begin{array}{c}
    \la \sigma^1_i,\sigma^2_j\ra=0\\ \la \sigma^1_i,\sigma^1_j\ra = 2\eta_{ij} =\la \sigma^2_i,\sigma^2_j\ra
    \end{array} \right) \\
    \left(\begin{array}{c}
    \la \sigma^1_i,\sigma^2_j\ra=0\\ \la \sigma^1_i,\sigma^1_j\ra = 2 \sqrt{\lambda}\, \eta_{ij} =-\la \sigma^2_i,\sigma^2_j\ra
    \end{array} \right)
    \end{array} \right.
\end{align}
Similarly to what was done in equation \eqref{Lagrangians I and II}, we can rewrite the 4d BF Lagrangian \eqref{4d BF Lagrangian} in the self-dual basis. We then get the two Lagrangians
\begin{subequations}
  \begin{align}
    L^+ &= 2 (L_1+L_2), \\
    L^- &= \f{2}{\sqrt{\lambda}} (L_1-L_2),
  \end{align}
\end{subequations}
which are built from a pair of uncoupled BF Lagrangians
\begin{equation}
  L_i =  B_i \wedge F_i - \kappa  B_i \wedge B_i.
\end{equation}

Now, the whole construction of section \ref{sec:4ddd?} can be applied to each $L_i$ independently of the other one. For each $L_i$ we have two usual gauge symmetries, associated to two charges
\begin{align}
    \slashed \delta \cJ_i(\alpha_i) = \oint_S \alpha_i\delta B_i,
    \q\q
    \slashed \delta \cT_i(\phi_i) = \oint_S \delta A_i \wedge \phi_i,
\end{align}
and a diffeomorphism symmetry which can be realized as a field-dependent combination of these gauge transformations. For a tangential vector field this diffeomorphism leads to the integrable charge
\begin{equation}
  \cD_i(\xi_i) = \cJ_i(\xi_i \ip A_i) + \cT_i(\xi_i \ip B_i).
\end{equation}
From the viewpoint of the initial BF Lagrangian, the algebra of quadratic charges is then naturally spanned by $\cD_{1,2}$, which form the algebra
\begin{subequations}
  \begin{align}
    \{\cD_1(\xi),\cD_1(\zeta)\} &=-\cD_1([\xi,\zeta]), \\
    \{\cD_2(\xi),\cD_2(\zeta)\} &=-\cD_2([\xi,\zeta]) ,\\
    \{\cD_1(\xi),\cD_2(\zeta)\} &=0,
  \end{align}
  \label{poidiffeo}
\end{subequations}
corresponding to two commuting copies of $\text{diff}(S^2)$. In order to recover the form \eqref{4d BF D* and D algebra} of the algebra, we need to go back to the Cartan basis by inverting the change of basis \eqref{eq:change_basis}. This is easily done by considering the combinations
\be
  \cD(\xi) = \cD_1(\xi) + \f{1}{\sqrt{\lambda}}\cD_2(\xi),\q\q \cD^{*}(\xi) = \cD_1(\xi) - \f{1}{\sqrt{\lambda}} \cD_2(\xi).
\ee

It is then immediate to generalize this construction. Indeed, the main point was that we were able to rewrite the BF Lagrangian in terms of two independent Lagrangians (which are also of the BF type in the present case). However, it is evident that there is no reason to limit the construction to a rewriting in terms of only two independent Lagrangians. For example, let us consider 4d BF with gauge group given by $\g = \su(2) \oplus \su(2) \oplus \su(2)$. The total connection $\cA$ and the 2-form $\cB$ can then be split into three components
\be
  \cA = A_1^i \sigma^1_i + A_2^i \sigma^2_i + A_3^i \sigma^3_i ,\q\q \cB = B_1^i \sigma^1_i + B_2^i \sigma^2_i + B_3^i \sigma^3_i,
\ee
where each $\sigma^a$ for $a=1,2,3$ generates $\su(2)$ and satisfies
\begin{equation}
  [\sigma_i^a,\sigma^b_j] =2 \delta^{ab} \epsilon_{ij}{}^k\sigma^{a}_k .
\end{equation}
Using this basis, the 4d BF Lagrangian for $\g = \su(2) \oplus \su(2) \oplus \su(2)$ can be rewritten into three independent Lagrangians $L_i$ for $\su(2)$ as we did previously. Each Lagrangian has its corresponding diffeomorphism charge, such that the full theory is equipped with three generators for the quadratic charges. Explicitly, the Lagrangian \eqref{4d BF Lagrangian} depends, as usual, on the choice of a Killing form. In this particular case, three independent Killing forms exist, which are given by
\begin{subequations}
  \begin{align}
    &\la \sigma^a_i,\sigma^b_j\ra_{+,+}=0 \;, \q\q \la \sigma^1_i,\sigma^1_j\ra_{+,+} = \la \sigma^2_i,\sigma^2_j\ra_{+,+} = \la \sigma^3_i,\sigma^3_j\ra_{+,+} = \eta_{ij} \;, \\
    &\la \sigma^a_i,\sigma^b_j\ra_{-,+}=0 \;,  \q\q \la \sigma^1_i,\sigma^1_j\ra_{-,+} = \la \sigma^2_i,\sigma^2_j\ra_{-,+} = \eta_{ij} = -\la \sigma^3_i,\sigma^3_j\ra_{-,+} \;, \\
    &\la \sigma^a_i,\sigma^b_j\ra_{+,-}=0 \;,  \q\q \la \sigma^1_i,\sigma^1_j\ra_{+,-} = \la \sigma^3_i,\sigma^3_j\ra_{+,-} = \eta_{ij} = -\la \sigma^2_i,\sigma^2_j\ra_{+,-} \;.
  \end{align}
\end{subequations}
Note that one can naturally construct a Killing form $\la\cdot\, , \cdot \ra_{-,-}$ where both the coupling for $\sigma_i^{2}$ and $\sigma_i^{3}$ have a $-1$ factor in terms of the above pairings. In terms of these (four) choices of pairings, the action \eqref{4d BF Lagrangian} takes the form
 \begin{align}\label{actionBFself11}
    L^{\epsilon_1,\epsilon_2} &= \la \cB \wedge \cF \ra_{\epsilon_1,\epsilon_2}  - \alpha \la \cB \wedge \cB \ra_{\epsilon_1,\epsilon_2}  = L_1+ \epsilon_1 L_2+ \epsilon_2 L_3, \q \epsilon_i=\pm1, \q i=1,2.
 \end{align}
 As explained previously, each Lagrangian $L_i$ then has its own diffeomorphism charge $\cD_i$, $i=1,2,3$, with the standard Poisson algebra for the diffeomorphisms. Since we have now three charges, the space spanned by the diffeomorphism charges is now three-dimensional from the viewpoint of the full theory. As above, instead of considering the basis $\cD_i$, we can consider the basis given in terms of the natural diffeomorphism associated to $L^{\epsilon_1,\epsilon_2}$, which is
 \begin{align}
     \cD^{\epsilon_1,\epsilon_2}(\xi) = \cD^1(\xi) + \epsilon_1 \cD^2(\xi) + \epsilon_2 \cD^3(\xi).
 \end{align}
Indeed, it is immediate to check that each diffeomorphism charge $\cD^{\epsilon_1,\epsilon_2}$ is the natural one associated to the Lagrangian $L^{\epsilon_1,\epsilon_2}$, while the other diffeomorphisms are the ``dual'' ones. The interpretation of these diffeomorphisms is a generalization of the previous case. In the sector 1, we always have a diffeomorphism along $\xi$, but in the sectors 2 and 3 we have diffeomorphisms directions $\epsilon_1 \xi$ and $\epsilon_2 \xi$ respectively. Indeed, looking at how $\cD^{\epsilon_1,\epsilon_2}$ acts on the fields we find
\bsub
\be
    \delta _\xi^{\cD^{\epsilon_1,\epsilon_2}}A_1 = \cL_\xi A_1, \q\q    \delta _\xi^{\cD^{\epsilon_1,\epsilon_2}}A_2 =\cL_{\epsilon_1\xi} A_2,\q\q \delta _\xi^{\cD^{\epsilon_1,\epsilon_2}}A_3 =\cL_{\epsilon_2\xi} A_3, \\
 \delta _\xi^{\cD^{\epsilon_1,\epsilon_2}}B_1 = \cL_\xi B_1, \q\q    \delta _\xi^{\cD^{\epsilon_1,\epsilon_2}}B_2 =\cL_{\epsilon_1\xi} B_2,\q\q \delta _\xi^{\cD^{\epsilon_1,\epsilon_2}}B_3 =\cL_{\epsilon_2\xi} B_3.
\ee
\esub
These diffeomorphisms form of course a closed algebra, given by
\be
     \{\cD^{\epsilon_1,\epsilon_2}(\xi),\cD^{\epsilon_3,\epsilon_4}(\zeta)\} = -\cD^{\epsilon_1\epsilon_3,\epsilon_2\epsilon_4}([\xi,\zeta]), \quad \epsilon_i=\pm1.
\ee
As we have shown in the main text, the existence of a self-dual basis is not necessary for the construction to work. It therefore begs the question of which conditions on the algebra are necessary for the existence of additional quadratic charges, and how many of them can be constructed. We are leaving this interesting question for future work.

\bibliographystyle{Biblio}
\bibliography{biblio}

\end{document}